\documentstyle[prl,aps]{revtex}
\begin{document}
%------------------------------------------------------------------------------
\title{Statistical mechanics of warm and cold unfolding in proteins}
\author{Alex Hansen, \footnote{Permanent address: Department of Physics,
Norwegian University of Science and Technology, NTNU, N--7034 Trondheim, 
Norway}\footnote{Electronic Address: Alex.Hansen@phys.ntnu.no}
Mogens H. Jensen,\footnote{Electronic Address: mhjensen@nbi.dk}
Kim Sneppen\footnote{Electronic Address: sneppen@nbi.dk} 
and Giovanni Zocchi\footnote{Electronic Address: zocchi@nbi.dk}}
\address{Niels Bohr Institute and NORDITA, Blegdamsvej 17, DK-2100 {\O}, 
Denmark}
\date{\today}
\maketitle
%------------------------------------------------------------------------------

\begin{abstract} 
We present a statistical mechanics treatment of
the stability of globular proteins
which takes explicitly into account the coupling between
the protein and water degrees of freedom.
This allows us to describe both the 
cold and the warm unfolding, thus qualitatively
reproducing the known thermodynamics of proteins.
\end{abstract} 
\vspace{0.5cm}
Classification: {\it Biophysics}
\vspace{1.0cm}

The folded conformation of globular proteins is a state of 
matter peculiar in more than one respect. The density is 
that of a condensed phase (solid or liquid), 
and the relative position of the atoms is, on average, fixed; 
these are the characteristics of the solid state. 
However, solids are either crystalline or amorphous,  
and proteins are neither: the folded structure, while ordered in the sense
that each molecule of a given species is folded in the same way, 
lacks the translational symmetry of a crystal. 
In Schr{\"o}dinger's words, proteins are
``aperiodic crystals". 
Unlike any other  
known solids, globular proteins are not really rigid, being able to perform  
large conformational motions while retaining locally 
the same folded structure.  
Finally, these are mesoscopic systems, consisting of a few thousand atoms. 

Quantitatively, the peculiarities of this state of matter are 
perhaps best appreciated from the thermodynamics. 
Delicate calorimetric measurements \cite{Priv1,Priv2,Privalov} 
on the folding transition of globular proteins reveals
the following picture: first the transition is first order,
at least in the case of single domain proteins.
Secondly, the stability of the folded state,  
i.e.\ the difference in Gibbs potential 
$\Delta G$ between the unfolded and the folded state is
at most a fraction of $k T_{room}$ per aminoacid. 
Following Privalov \cite{Privalov}, we will refer to this property as 
``cooperativity". The Gibbs potential difference $\Delta G$, 
as a function of temperature, is non monotonic: 
it has a maximum around room temperature (where $\Delta G > 0$  
and so the folded form is stable), then crosses zero and 
becomes negative both for higher and lower temperatures. 
Correspondingly, the protein unfolds not only at high, 
but also at low temperatures. This 
phenomenon of ``cold unfolding", which is observed experimentally, 
is most peculiar: solids usually do not melt upon cooling! 
For temperatures around the cold unfolding transition and below, the 
enthalpy difference $\Delta H$ between the unfolded and the 
folded state is negative; this means that cold unfolding 
proceeds with a release of heat (a negative latent heat), 
as is also observed experimentally; at the higher unfolding 
transition, on the contrary, $\Delta H > 0 $ which corresponds 
to the usual situation of a positive latent heat. 
Fig.1 shows Privalovs measurements 
of the specific heat of myoglobin \cite{Privalov}.
There are two peaks in the specific heat, corresponding to the 
two unfolding transitions, and a large gap $\Delta C$ in the 
specific heat between the unfolded and the folded state.  
This gap is again peculiar to proteins: usually, 
for a melting transition $\Delta C \approx 0 $  
(e.g. for ice at 0 C $~ C = 1.01 ~ cal/g K$ while for water at 0 C $~  
C = 1.00 ~ cal/g K$). The existence of this gap $\Delta C $ is related to 
the phenomenon of cold unfolding \cite{Privalov}. 
 
From the microscopic point of view, 
the main driving force for folding is the  
hydrophobic effect. In the native state of 
globular proteins hydrophobic residues  
are generally found on the inside of the molecule,  
where they are shielded from the water, while hydrophilic residues  
are typically on the surface. In the following we refer to the 
difference in free energy between hydrophobic residues interacting 
with each other in the core of the folded protein and these same 
residues interacting with the water in the unfolded structure, 
including any changes in the microscopic states of the water, 
as the ``hydrophobic interaction". 
Hydrogen bonds within the regular elements of secondary 
structure ($\alpha$ helices and $\beta$ sheets), while 
necessary for the stability of the native state, can hardly 
be thought of as providing the positive $\Delta G$ of the 
folded structure, since the unfolded structure would form 
just as many hydrogen bonds with the water. When the 
protein unfolds, the hydrophobic residues of the interior 
are exposed; this accounts for most of the gap in the 
specific heat $\Delta C$ \cite{Privalov}, according to 
the known effect that dissolving hydrophobic substances 
in water raises the heat capacity of the solution \cite{Edsall}. 

As in other branches of physics, once the thermodynamics 
of a system is known it is desirable to develop a corresponding 
statistical mechanics picture.  
Several models have been proposed which address some aspects of the  
folding transition. In the ``zipper model" \cite{zipper} , which 
was introduced to describe the helix - coil transition, the relevant 
degrees of freedom (conformational angles) are treated as a set of variables 
which can take two values: one corresponding to matching the ordered structure 
(helix), and the other corresponding to the ``coil" state. The problem 
is then equivalent to the 1-D Ising model.  
A related parametrization for the 3-d folding transition
has been proposed by Zwanzig \cite{Z95}, describing the folding
transition in terms of variables 
which each awards match with the correct ground state.
A zipper model that deals with the initial pathway of protein
folding has been proposed by Dill, Fiebig and Chan \cite{D93}.
For a review see \cite{Dill95}.
A recent discussion of hydrophobicity in protein 
folding is in ref.\ \cite{Tang}.

However, to our knowledge no model exists which reproduces 
all the thermodynamic features surveyed above. With the present 
work, we address this question. 

We start with a Hamiltonian which we have recently 
introduced \cite{us} to describe self-assembly of a 
cooperatively stabilized (in the sense defined above) 
structure:  
\begin{equation}
\label{nat1}
H=-\ (\varphi_1+\varphi_1\varphi_2+
 \varphi_1\varphi_2\varphi_3+\cdots +
 \varphi_1\varphi_2\cdots\varphi_N)~~~,
\end{equation}
here the $\varphi$'s are variables which take on the values 
0 and 1, and, in the spirit of the zipper model, we define 
the ground state ($\varphi_i = 1 ~ \forall ~ i$ ) as the template 
corresponding to the ordered, aperiodic structure, i.e. the folded 
state. One can think of the $\varphi$'s as appropriately coarse-grained 
angle variables which define the conformation of the polypeptide chain.
The above Hamiltonian is then a description
of a system that has a specific folding pathway; 
a property that is well documented for proteins 
\cite{Levinthal,Fersht,Creighton,Creightonbook,Z96}.
As discussed in \cite{us}, this system has
a first order phase transition from an ordered 
to a disordered state at temperature $T_m = 1/\ln 2$.
The Hamiltonian (\ref{nat1}) exhibits a hierarchical structure: 
unless $\varphi_1,\varphi_2,\cdots,\varphi_i = 1$ , 
it does not matter what value the rest of the variables $\varphi_{i+1},
\varphi_{i+2},\cdots,\varphi_N$ assume. As a consequence,  
the system displays cooperativity, in the sense that the 
binding energy per degree of freedom in the ordered state, 
for $T \approx T_m$, is only of order $k T_m$. 

In order to proceed further, 
it is necessary to take also the water into account.
The relevant physics here is that dissolving a hydrophobic 
substance in water causes a large decrease in the entropy 
of the system \cite{Privalov}. This entropy change is 
attributed to a partial ordering of the water molecules 
around the hydrophobic solute. 
The gradual melting of this additional structure upon 
heating causes the increase in heat capacity. 
Consequently, we introduce a second set of variables 
$\mu_1, \mu_2, ... , \mu_N$ which describe the water. 
These water degrees of freedom couple to 
the unfolded protein degrees of freedom 
because these expose hydrophobic amino 
acids to the water.
This is achieved by the Hamiltonian:
\begin{equation}
\label{eq2}
H  =  - (\varphi_1+\varphi_1\varphi_2+\varphi_1\varphi_2\varphi_3+\cdots +
 \varphi_1\varphi_2\cdots\varphi_N) \nonumber  
  -  [ (1-\varphi_1)\mu_1 + (1-\varphi_1\varphi_2)\mu_2 + \cdots + 
(1-\varphi_1\varphi_2\cdots\varphi_N)\mu_N ]   
\end{equation}

The $\varphi$'s take on the values 0 or 1, as before. 
Each of the $\mu_i$'s can take a value from the set 
${\cal E}_{min} + s \Delta {\cal E}$,  $s = 0,1,2,\cdots,g-1 \}$
where  ${\cal E}_{min} < 0$ , $\Delta {\cal E} > 0$. 
If at least one of the variables $\varphi_1 , ... , \varphi_i$  
equals zero, the corresponding contribution of the $i$'th 
water variable to the energy is $\mu_i$ and zero otherwise.
Therefore when $\varphi_1 \cdots \varphi_i=1$, the states
for the corresponding water degree of freedom $\mu_i$ 
are degenerate with zero contribution to the 
energy and degeneracy $g$. 

To reiterate, the physical meaning of this Hamiltonian is that 
the water molecules in contact with an unfolded portion of the protein 
go to a lower entropy state (compared to the water molecules in contact 
with a folded portion), but also, for low temperatures, to a more
tightly bound state. The more specific features of the model (2),  
e.g. the structure of the energy spectrum, the particular coupling 
of the $\mu$'s to the $\varphi$'s, etc. can be varied while maintaining 
the overall thermodynamic behavior described below. Here we just present 
the case which is simplest to solve analytically. 

The calculation of the partition function is straightforward.
We parametrize the states of the system
by the number $n$ of consecutive matches
$\varphi_1 =1, \varphi_2=1,..., \; \varphi_{n}=1$ 
and ending with $\varphi_{n+1}=0$ 
and the values $\{s_{n+1},... ,s_N\}$ where each
$s_i \in \{0,1,2,..., g-1\}$ for the $(N-n)~$ $\mu$ 
variables coupled to the unfolded portion of the protein.
The energy of this state is
\begin{equation}
\varepsilon (n,s_{n+1},..., s_N )
~=~ - n~ {\cal E}_0  ~+~ \sum_{i = n+1}^N  
({\cal E}_{min} + \Delta {\cal E}~ s_i)
\end{equation} 
where we have introduced the energy scale ${\cal E}_0$ for the protein 
variable in order to 
make the formulas dimensionally more transparent
(up to now we used ${\cal E}_0=1$). 
Denoting $\beta=1/T$ as the reciprocal temperature,
the partition function is 
\begin{equation}
Z = \sum_{n=0}^{N-1} \; 
2^{N-n-1} g^{n} \;
\sum_{s_{n+1}=0}^{g-1}  
\sum_{s_{n+2}=0}^{g-1} \cdots \sum_{s_N=0}^{g-1} 
\exp(\;- \beta \varepsilon (n,s_1,\cdots, s_N ) ) 
~~+~ g^N \exp( \;\beta {\cal E}_0 N) 
\end{equation} 
In the above equation  the factor $2^{N-n-1}$ is the 
degeneracy of the unfolded protein degrees of freedom and
the factor $g^{n}$ is the degeneracy of water which is not
exposed to the inside of the protein. Factorizing the sums 
over $s_i$ into partition functions $Z_w$ 
for each water degree of freedom we write:
\begin{equation}
\label{zwater}
Z \;=\; \frac{1}{2} ~ (2 Z_w )^{N} \sum_{n=0}^{N-1} 
\left( \frac{ g \exp(\beta {\cal E}_0) }{2 Z_w} \right)^{n}  
~~+~ \left( g \exp( \;\beta {\cal E}_0 ) \right)^N
\end{equation}
where the phase space for a water degree of freedom exposed
to an unfolded protein degree of freedom is 
\begin{equation}
Z_w \; =\; \sum_{s=0}^{g-1} \exp( -\beta ( {\cal E}_{min} + 
s \Delta {\cal E} ) )
\; =\; \frac{(\exp(-\beta {\cal E}_{min}) - 
\exp(-\beta {\cal E}_{max})}{
( 1 - \exp(-\beta \Delta {\cal E}))} 
\end{equation}
where ${\cal E}_{max}={\cal E}_{min} + g \Delta {\cal E}$.
From Eq.\  (\ref{zwater})one sees directly that the state of the system 
is determined by the size of the quantity
\begin{equation}
\frac{ g \exp(\beta {\cal E}_0)}{2 Z_w} \;=\; \exp(\beta \Delta f) 
\end{equation}
If $\Delta f > 0$ then  the system will be in the folded state
because the sum in Eq.\ (\ref{zwater}) is dominated by the last term, whereas
for $\Delta f<0$ the system will be unfolded.

The sum in Eq.\ (\ref{zwater}) can be readily performed
and the total partition function is 
\begin{equation}
Z \;=\; \frac{1}{2} (2 Z_w )^{N} ~ \frac{
1 - \left(  g \exp({\cal E}_0 \beta)~ / ~(2 Z_w) \right)^{N} 
}{
1 - \left(  g \exp({\cal E}_0 \beta)~ / ~(2 Z_w) \right) 
}
~~+~ \left( g \exp( \;\beta {\cal E}_0 ) \right)^N
\end{equation}
The free energy is $F~=~-T \ln (Z)$, 
the energy $E~=~-d \ln(Z)/d\beta$ and the heat capacity
$ C~ =~ dE/dT$. 
Because there is no pressure in the model, 
the energy $E$ takes the place of 
the enthalpy $H = E + p V$ and the free energy 
$F = E - T S$ takes the place of 
the Gibbs potential $G = H - T S$. 
In Fig.\ref{fig2} we show 
the heat capacity per degree of freedom for four different
choices of ${\cal E}_{min}$, representing four different
values of the chemical potential, which we discuss later.
The characteristic feature is that there are two peaks
corresponding to warm and cold unfolding, and a gap
$\Delta C$ in the heat capacity between the unfolded and the
folded form. At higher temperatures, i.e. $T > g \Delta {\cal E}$,
the gap goes to zero because the water becomes effectively 
degenerate again. In Fig.3a we show the order parameter
$\langle n \rangle$ as function of temperature.
The figure indeed confirms that the protein is folded between
the two transitions.

We now calculate explicitly the difference in the thermodynamic 
functions between the unfolded and the folded state. 
We consider these quantities per degree of freedom.
The thermodynamic functions associated to a folded (f) protein variable
is the energy $e_f=-{\cal E}_0$, the entropy $s_f=\ln(g)$
and the free energy $f_f=-{\cal E}_0-T\ln(g)$.
The free energy associated to an unfolded (u) protein variable
is given by the corresponding partition function of water
multiplied by the degeneracy factor of
an unfolded part of the protein: $f_u=-T\ln(Z_w ~ 2)$.
The difference in free energy between folded and unfolded 
state is accordingly
\begin{equation}
\Delta f ~=~ f_u ~-~ f_f ~=~ 
T~ \ln( \frac{ g~ \exp( \beta {\cal E}_0)}{2 ~Z_w} )
\end{equation}
which is the quantity we earlier identified as the one 
which decides whether the system cooperatively selects 
the folded or the unfolded state.
To get a simple insight in this formula we rewrite it 
for small energy level spacings $\Delta {\cal E} <<T$:
\begin{equation}
\label{deltaf}
\Delta f ~=~ 
{\cal E}_0 ~+~ {\cal E}_{min} ~+~ 
T \ln ( \frac{g \Delta {\cal E}}{2 ~ T} )
~-~T \ln \left( 1- \exp(-({\cal E}_{max} -{\cal E}_{min})/T) \right)
\end{equation}
From this expression for the difference in free energy one easily obtains the
corresponding differences in energy, entropy and specific heat. In
particular, we obtain a gap in the specific heat 
between the folded and unfolded state
$\Delta c =  ( \Delta {\cal E} / T)^2 / (e^{\Delta {\cal E} / T} - 1)^2 ~ 
e^{\Delta {\cal E} / T} ~\sim~ exp( \Delta {\cal E} / T) \sim 1$ 
for temperatures $T \in [\Delta {\cal E},{\cal E}_{min}+{\cal E}_{max}]$, 
see Fig.\ref{fig2}.

To simplify the discussion let us consider the limit of large
${\cal E}_{max}$ in (\ref{deltaf}). It is easily seen that $\Delta f$
has a maximum at the temperature
$T_m \approx g \Delta {\cal E} / 2 e$ . 
The corresponding value of $\Delta f$ is 
$ \Delta F(T_m) \approx ~ ({\cal E}_{min} + 
{\cal E}_0) ~ + ~ g \Delta {\cal E} / 2 e $ , 
so the condition for the existence of a region of 
stability of the ordered structure ($\Delta f > 0$) is: 
 
\begin{equation} 
\frac{g \Delta {\cal E}}{2 e} > - ({\cal E}_{min} + {\cal E}_0) . 
\end{equation} 
This is of course always satisfied if 
$ ({\cal E}_{min} + {\cal E}_0) > 0$ , however 
the more interesting situation is 
$ ({\cal E}_{min} + {\cal E}_0) < 0$, since then 
$\Delta F < 0$ at sufficiently low temperature, 
i.e. the phenomenon of cold unfolding 
appears. Under these conditions $\Delta E$ is 
also negative at sufficiently low temperature 
which means that we have a negative 
latent heat for cold unfolding. Fig.3b
shows these thermodynamic functions. They qualitatively 
reproduce the known thermodynamic behavior of globular proteins 
as described in the introduction \cite{Privalov}.  
In the present description the mechanism for the transitions is
the following.  
At high temperature the entropy gain of the protein
causes the unfolding. As temperature is lowered the water exposed
to hydrophobic parts of the protein gets more and more ordered,
and consequently the system gains more
entropy by shielding the hydrophobic residues from the water (folding).
As the temperature is lowered even further
the cold unfolding transition occurs: here
all entropy contributions to free energy are small
and the dominating effect is the coupling energy
between the water and the unfolded protein.

Coming back to the partition function (3) and  (4), we may write: 
\begin{equation} 
{\cal E} = - N {\cal E}_0 ~+~ (N-n) ({\cal E}_0 + {\cal E}_{min}) 
+ \sum_{i=n + 1}^N ~ \Delta {\cal E}~ s_i ~=~ - N {\cal E}_0 ~+~
\sum_{i=n + 1}^N ~[\Delta {\cal E}~ s_i + {\cal E}_0 + {\cal E}_{min}] 
\end{equation} 

and 

\begin{equation} 
Z =  e^{\beta N {\cal E}_0}  \sum_{n=0}^{N-1} ~  
2^{N-n-1} g^{n} \;
\sum_{\lbrace s_i \rbrace} ~ 
e^{- \beta \sum_{i=n + 1}^N ~ ({\cal E}_i - \mu )} ~
~~+~ g^N \exp( \;\beta {\cal E}_0 N) 
\end{equation} 
where we have set 
${\cal E}_i = \Delta {\cal E} ~ s_i ~, 
~ \mu = - ( {\cal E}_0 + {\cal E}_{min})$. 
From this expression for $Z$ we can identify $\mu$ with the 
chemical potential \it{ of the water } \rm , or, to be more 
precise, the difference in chemical potential of the water 
when it is in contact with the hydrophobic 
interior of the protein and when it is not. 
Therefore, $\mu > 0$ is the physically 
relevant situation. Experimentally, $\mu$ can be changed by adding denaturants, 
changing pH, etc., which indeed alters the stability of the ordered structure. 
The four curves in Fig.\ref{fig2}, 
which are to be compared with the experimental
data in Fig.\ref{fig1}, are the results of 
the model for different values
of the chemical potential $\mu$. 

In conclusion, this paper introduces a new model for the stability
of proteins which reproduces the known thermodynamics.
We obtain: 1. first order unfolding transitions; 2. both warm and
cold unfolding; 3. cooperativity in the sense that the free energy
difference stabilizing the folded state 
is only a fraction of $kT_{room}$ per degree of freedom; 4. a qualitatively
correct behavior of the specific heat both as a function of
temperature and chemical potential; 5. a gap in the specific heat between
the unfolded and folded state; 6. a negative
latent heat for the cold unfolding. 
A deficiency of the model is that our description of the water-protein
coupling is simplified. As a result, the two transitions
are too far apart in absolute temperature and in the model
the cold unfolding appears
sharper than the warm unfolding, which is not seen in experiment. 
This deficiency calls for some modifications, in particular by
introducing both hydrophobic and hydrophilic $\varphi_i$'s
one can influence the relative strength of the two transitions.

%--------------------------------------------------------------------
% BIBLIOGRAPHY
% --------------------------------------------------------------------

% --------------------------------------------------------------------
% FIGURE CAPTIONS
% --------------------------------------------------------------------
\begin{figure}
\caption[x]{
Calorimetric measurements of the specific heat of Myoglobin
at four different values of pH, as presented by Privalov in
ref. \cite{Privalov}.
At sufficiently low pH the native structure of the protein
never becomes stable, thus the protein
remains in its unfolded structure
with approximately constant heat capacity over the measured
temperature range. By increasing pH the native structure 
becomes stabilized for
intermediate temperatures, defining a transition to an 
unfolded state at both low and high temperatures, 
denoted respectively cold and warm denaturation.
There is a gap in the specific heat between the folded
and the unfolded states.
\label{fig1}
}
\end{figure}
\begin{figure}
\caption[x]{
Specific heat as function of temperature $T$ 
for the model,
with four different values of the chemical potential 
$\mu=-({\cal E}_0+{\cal E}_{min})=-1.0,~-1.1,~-1.2,~-1.5$
and with fixed level spacing $\Delta {\cal E}=0.2$,
$g=35$ and system size $N=60$.
As the chemical potential is lowered,
it becomes increasingly difficult to fold, and finally
for sufficiently low $\mu$ the protein 
stays unfolded.
\label{fig2}
}
\end{figure}
\begin{figure}
\caption[x]{
Average fraction of folded protein variables 
as function of temperature for 
$\mu=-({\cal E}_0+{\cal E}_{min})=-1.0$ and the 
other parameters as in figure 1.
The figure shows that in between the transitions the protein
is folded.\\

Fig. 3b. Difference of thermodynamic functions between 
folded and unfolded configurations for the same chemical
potential as in a). The
difference in free energy $\Delta f$ has a maximum
and becomes negative for both high and low temperature
(cold unfolding).
$\Delta S/N$ and $\Delta E/N$ increase with temperature.
$\Delta E/N $ is negative at the cold unfolding transition
corresponding to a negative latent heat.
\label{fig3a}
}
\end{figure}
% --------------------------------------------------------------------
\end{document}